\documentstyle[epsf]{article}          
\def\vec#1{\mbox{\boldmath $#1$}}

\catcode`@=11

\@addtoreset{equation}{section}
\catcode`@=12

\begin{document}

\title{{\small\centerline{\parbox[t]{29mm}{SINP/TNP/98-08 
IISc-CTS/2/1998}
\hfill February 1998 \hfill 
hep-th/9803009}} 
\bigskip \bigskip 
{\bf Functional derivation of Casimir energy \\ \bf at non-zero 
temperature}} 

\author{\bf Avijit K. Ganguly\\ 
\normalsize Centre for Theoretical Studies, Indian Institute of
Science, Bangalore 560012, INDIA\\  \smallskip\\ 
\bf Palash B. Pal\\ 
\normalsize Saha Institute of Nuclear Physics, 1/AF Bidhan-Nagar, 
	Calcutta 700064, INDIA}

\date{}
\maketitle 

\begin{abstract} \normalsize\noindent
Performing functional integration of the free Lagrangian, we find the
vacuum energy of a field. The functional integration is performed in a
way which easily generalizes to systems at non-zero temperature. We
use this technique to obtain the Casimir energy density and pressure
at arbitrary temperatures. 
\end{abstract}
\bigskip\bigskip

\section{Introduction}
If we have electromagnetic field in a restricted region, the vacuum
energy of the electromagnetic field shows up as an effective zero
point energy, which is known as Casimir energy. For a rectangular
region bounded by two perfectly conducting plates separated by a
distance $\zeta$ along the $z$-axis, the magnitude of this energy is
given by
	\begin{eqnarray}
E = - {\pi^2 L_xL_y \over 720 \zeta^3} \,,
	\end{eqnarray}
where $L_x$ and $L_y$ are the dimensions of the plates (assumed large)
along the other two axes. The energy density in the region between the
plates is therefore given by
	\begin{eqnarray}
\rho = - {\pi^2 \over 720 \zeta^4} \,.
\label{rhoresult}
	\end{eqnarray}

There are various ways that the magnitude of this energy density can
be derived. Here, we outline a method using functional integration of
the Lagrangian. We will show that this method easily extends to the
case when the region between the two plates is at a finite
temperature. Using this extension, we will calculate the Casimir
energy at finite temperature.

It is well known that the functional formulation of the
electromagnetic field is complicated because of gauge invariance. The
gauge volume has to be divided out by a Fadeev-Popov procedure. In
order to avoid these complications, we perform all our calculation
using a complex scalar field of mass $M$. In other words, we will
calculate the zero-point energy of a complex scalar field in the
specified geometry, assuming that this field satisfies the same
boundary conditions as the electromagnetic field at the boundaries of
the region. Once this is done, the result can be applied for the
electromagnetic field as well, since it has the same number of degrees
of freedom. We only need to put the mass of the field to be equal to
zero.

\section{The functional formulation}
Let us then start from the Lagrangian of a free scalar field:
	\begin{eqnarray}
{\cal L} = (\partial^\mu \phi^\dagger) (\partial_\mu \phi) -
M^2 \phi^\dagger \phi \,.
	\end{eqnarray}
The generating functional obtained from this Lagrangian is given by 
	\begin{eqnarray}
{\cal Z} = \int [{\cal D}\phi][{\cal D}\phi^\dagger] \; \exp \left( i
\int d^4x \, {\cal L} \right) \,.
	\end{eqnarray}
Since the free Lagrangian is quadratic, this integration can be
performed formally, and one obtains
	\begin{eqnarray}
{\cal Z} = {\rm det} \, \left( \partial^2 + M^2 \right)^{-1} \,.
	\end{eqnarray}
Thus, 
	\begin{eqnarray}
\ln {\cal Z} &=& -\, {\rm tr} \, \ln \left( \partial^2 + M^2
\right)  \,, 
	\end{eqnarray}
using a well-known identity that for any operator, the logarithm of
its determinant is the same as the trace of its logarithm.

The trace mentioned above runs not only over any internal degrees of
freedom that may be present (in the particular case at hand, there
isn't any), but also over the space-time points. Thus, more
explicitly, we can write
	\begin{eqnarray}
\ln {\cal Z} &=& -\, \int d^4x\; 
\left< x \left| \ln \left( \partial^2 + M^2 \right) \right| x \right>
\,. 
	\end{eqnarray}
Going over to the momentum representation, this can be written as
	\begin{eqnarray}
\ln {\cal Z} &=& -\, \int {d^4k \over (2\pi)^4} \; 
\ln \left( -k^2 + M^2 \right) \nonumber\\* 
&=& \int {d^4k \over (2\pi)^4} \int dM^2 \; 
{1 \over k^2 - M^2} \,,
\label{Sinf}
	\end{eqnarray}
where the integration over $M^2$ is indefinite. 
The integral on the right side is, of course, infinite. This is no
surprise. 
For a general system, the quantity $\cal Z$ can be interpreted as 
	\begin{eqnarray}
\ln {\cal Z} = -\, i f_{\rm vac} \,,
	\end{eqnarray}
where $f_{\rm vac}$ is the free energy density of the vacuum, so that
we can write 
	\begin{eqnarray}
f_{\rm vac} = i \int dM^2 \int {d^4k \over (2\pi)^4} \;
{1 \over k^2 - M^2} \,. 
\label{fvac}
	\end{eqnarray}
For a system at zero temperature, the free energy density is equal to
the internal energy density, $\rho_{\rm vac}$. The infinity obtained
for the integral on the right side then just corresponds to the well
known result in quantum field theory that the vacuum energy of a field
is formally infinite. Once normal ordered fields are employed, this
infinite contribution vanishes. Effectively, this amounts to setting
the zero of energy at the vacuum energy of the field. 

In the derivation above, we have implicitly assumed that the
space-time region is infinite in all directions. For a more general
situation, let us denote the obvious
generalization of Eq.~(\ref{fvac}) by
	\begin{eqnarray}
f _{\rm vac} = i\int dM^2 \int {[d^4k] \over
(2\pi)^4} \; {1 \over k^2 - M^2} \,,
\label{Sgeneral}
	\end{eqnarray}
where the square brackets around the momentum integration measure now
indicate that the measure has to be taken as appropriate in
a particular situation. In fact, if the momentum values happen to be
discrete, we should sum rather than integrate over the momenta.

In order to calculate the vacuum energy at a non-zero temperature
$1/\beta$, we can employ the imaginary time formalism, where the $k_0$
values for a bosonic field are quantized in the form~\cite{TFT}
	\begin{eqnarray}
k_0 = 2\pi in/\beta \,, 
	\end{eqnarray}
for arbitrary integers $n$. In this case, the integration over $k_0$
is to be replaced by a sum over the imaginary energies. Thus,
	\begin{eqnarray}
\int {[d^4k] \over (2\pi)^4} = 
{i\over \beta} \sum_{n=-\infty}^\infty \int {[d^3k] \over
(2\pi)^3}\,.
	\end{eqnarray}
We are still keeping the square brackets around the integration
measure for the spatial components of momentum in view of the fact
that in a restricted geometry as is relevant for the Casimir problem,
the spatial components of momenta will also not be continuous. 
Thus, for a scalar field at finite temperature, we can write 
	\begin{eqnarray}
f_\beta &=& - \, {1\over\beta}
\sum_{n=-\infty}^\infty 
\int {[d^3k] \over (2\pi)^3} \; \int dM^2 \; {1 \over (2\pi
in/\beta)^2 - \vec k^2 - M^2} \,. 
\label{Sbeta}
	\end{eqnarray}
The sum over $n$ present in
Eq.~(\ref{Sbeta}) can be performed by using the identity
	\begin{eqnarray}
\sum_{l=-\infty}^\infty {1\over l^2 + a^2} = {\pi\over a} \coth \pi a 
\,. 
\label{sum}
	\end{eqnarray}
Writing
	\begin{eqnarray}
\coth z = 1 + {2 \over e^{2z} -1} \,,
	\end{eqnarray}
we can rewrite Eq.~(\ref{Sbeta}) as
	\begin{eqnarray}
f_\beta &=& \int {[d^3k] \over (2\pi)^3} \; \int dM^2 \; 
{1 \over 2\omega_k} 
\left( 1 + {2 \over e^{\beta \omega_k} - 1 } \right) \,,
\label{Stotal}
	\end{eqnarray}
using the shorthand
	\begin{eqnarray}
\omega_k \equiv \sqrt{ \vec k^2 + M^2} \,.
\label{omegak}
	\end{eqnarray}

The integral over $M^2$ present in the resulting form is an indefinite
integral. Instead of using $\vec k$ and $M$ as independent parameters,
we can use $\vec k$ and $\omega_k$. This change can be effected by
making the substitution
	\begin{eqnarray}
dM^2 = 2\omega_k \, d\omega_k \,,
	\end{eqnarray}
which gives
	\begin{eqnarray}
f_\beta &=&  \int {[d^3k] \over (2\pi)^3} \; \int d\omega_k \; 
\left( 1 + {2 \over e^{\beta \omega_k} - 1 } \right) \equiv f_\infty +
f'_\beta\,.
\label{2terms}
	\end{eqnarray}
Clearly, the first term in the integrand gives a temperature
independent part, which is the contribution at zero temperature (i.e.,
at $\beta\to\infty)$. As indicated, this term will be denoted by
$f_\infty$. The second term is the temperature dependent contribution,
which we have denoted by $f'_\beta$. We now discuss these
contributions one by one.

\section{The temperature independent contribution}
{}For the first term, which we denoted by $f_\infty$, we
can straightaway perform the indefinite integral over $\omega_k$ to
write 
	\begin{eqnarray}
f_\infty &=& \int {[d^3k] \over (2\pi)^3} \; \omega_k 
\label{1stterm}
	\end{eqnarray}
To proceed, we now use the identity
	\begin{eqnarray}
\omega_k &=& {1 \over \Gamma(-1/2)} \int_0^\infty {ds \over s^{3/2}} \;
e^{-s\omega_k^2} \nonumber\\*
&=& - \, {1 \over \sqrt{4\pi}} \int_0^\infty {ds \over s^{3/2}} \;
\exp \left[ -s (\vec k^2 + M^2) \right] \,.
\label{omega}
	\end{eqnarray}
Substituting this form into the first term of Eq.~(\ref{2terms}), we
can perform the $\vec k$ integration if the space were infinite. This
would give
	\begin{eqnarray}
f_\infty = -\, \int_0^\infty {ds \over s} \; {1 \over (4\pi s)^2} 
e^{-s M^2} \,.
\label{finf}
	\end{eqnarray}
This is essentially the form obtained by Schwinger \cite{schwinger}
through his proper-time formalism. 
If the integration is performed now, we would get a result
proportional to $\Gamma(-2)$, which would be infinite, as remarked
earlier. 

But we also said that this energy is not really relevant for us. In
the Casimir geometry, i.e., in the region between two infinite
conducting plates at $z=0$ and $z=\zeta$, the component of momentum
perpendicular to the plates will be quantized. For an electromagnetic
field, these quantized values will be given by
	\begin{eqnarray}
k_z = {\pi l \over \zeta} 
	\end{eqnarray}
with arbitrary integers $l$ so that the potential can vanish at both
the plates. As we said in the Introduction, we will take this same
boundary condition for the complex scalar field as well. Thus, for the
Casimir geometry, we should write
	\begin{eqnarray}
\int {[d^3k] \over (2\pi)^3} = 
{1\over 2\zeta} \sum_{l=-\infty}^\infty \int {d^2k_\perp \over
(2\pi)^2}\,,
\label{d3k}
	\end{eqnarray}
where $k_\perp$ indicates momentum values in the plane perpendicular
to the $z$-axis. 
Using Eqs. (\ref{1stterm}) and (\ref{omega}) as before, we can now
perform the integration over $k_\perp$ to obtain
	\begin{eqnarray}
f_\infty &=& -\, \int_0^\infty {ds \over s} \; {1 \over (4\pi s)^{3/2}} \,
e^{-s M^2} \; {1\over 2\zeta} \sum_{l=-\infty}^\infty \exp \left( -\,
{\pi^2 s \over \zeta^2}\, l^2 \right) \nonumber\\*
&=& -\, \int_0^\infty {ds \over s} \; {1 \over (4\pi s)^{3/2}} \,
e^{-s M^2} \; {1\over 2\zeta} \, \vartheta_3 
\left( 0, {i\pi s \over \zeta^2} \right) \,,
\label{zonetheta}
	\end{eqnarray}
using the standard definition of the Jacobi
$\vartheta$-function\footnote{See, e.g., Ref.~\cite{GR}, \S\,8.180,
eq.~4 and the comment below.}: 
	\begin{eqnarray}
\vartheta (u,\tau) = \sum_{l=-\infty}^\infty \exp \left( 2lui + 
i\pi\tau l^2 \right) \,.
	\end{eqnarray}

The expression for $f_\infty$ given above can be simplified, noting
that mathematically, the problem of Casimir geometry at zero
temperature is equivalent \cite{toms} to the problem of infinite
geometry at non-zero temperature, tackled through the imaginary time
formalism. We therefore follow a procedure enumerated by Dittrich
\cite{dittrich} for non-zero temperatures, adapting it to the present
case. For this, one notes the following property of the Jacobi
$\vartheta$-function:
	\begin{eqnarray}
\vartheta (0,i\tau) = {1\over \sqrt\tau} \vartheta (0, {i\over \tau})
\,,
\label{thetaidentity}
	\end{eqnarray}
which enables us to write
	\begin{eqnarray}
{1\over 2\zeta} \vartheta_3 
\left( 0, {i\pi s \over \zeta^2} \right) 
&=& {1\over \sqrt{4\pi s}} \vartheta_3 
\left( 0, {i\zeta^2 \over \pi s} \right) \nonumber\\* 
&=& {1\over \sqrt{4\pi s}} \sum_{l=-\infty}^\infty \exp \left( -\,
{\zeta^2 l^2 / s} \right) \,.
	\end{eqnarray}
Putting this form back in Eq.~(\ref{zonetheta}), we obtain
	\begin{eqnarray}
f_\infty &=& -\, \int_0^\infty {ds \over s} \; {1 \over (4\pi s)^2} \,
\sum_{l=-\infty}^\infty \exp \left( -\,
{\zeta^2 l^2 \over s} - sM^2 \right) \,.
	\end{eqnarray}
There is a summation over $l$. Notice that, the $l=0$ term in this sum
is exactly same as the result obtained in Eq.~(\ref{finf}) for
infinite space. Thus, this 
result is also infinite. However, the physically important quantity is
the difference of this quantity and the corresponding result for the
infinite space, which is:
	\begin{eqnarray}
f'_\infty \equiv 
f_\infty - f_{\rm vac} = -
2 \int_0^\infty {ds \over s} \; {1 \over (4\pi s)^2} \,
\sum_{l=1}^\infty \exp \left( -\,
{\zeta^2 l^2 \over s} - sM^2 \right) \,.
\label{0temp}
	\end{eqnarray}
The integral over $s$ can be expressed in terms of a modified Bessel
function, defined by\footnote{See, e.g., Ref.~\cite{GR}, \S\,8.432,
eq.~6.} 
	\begin{eqnarray}
K_\nu(z) = {1\over 2} \left( {z\over 2} \right)^\nu \int_0^\infty {dt
\over t^{\nu+1}} \; \exp \left( -t - {z^2 \over 4t} \right) \,.
\label{modBessel}
	\end{eqnarray}
Writing $sM^2=t$ in Eq.~(\ref{0temp}), the expression can be rewritten
in the form
	\begin{eqnarray}
f'_\infty = -\,
{2M^4 \over (4\pi)^2} \, \sum_{l=1}^\infty 
\int_0^\infty {dt \over t^3} 
\exp \left( -\,
{\zeta^2 l^2 M^2 \over t} - t \right) = 
-\, {M^2 \over 4\pi^2 \zeta^2} \, \sum_{l=1}^\infty {1\over l^2}
K_2(2\zeta lM) \,.
	\end{eqnarray}
This result is twice that of Ambjorn and Wolfram \cite{AW}, who
calculated the same quantity for a real scalar field.

For the $M\to 0$ limit which is relevant for finding the result for
the electromagnetic field, we can use the form 
	\begin{eqnarray}
K_2(z) \approx 2z^{-2} \,,
\label{K2smallz}
	\end{eqnarray}
which is valid for small $z$. This gives
	\begin{eqnarray}
f'_\infty = 
-\, {1\over 8\pi^2 \zeta^4} \, \sum_{l=1}^\infty {1\over l^4} = -\,
{\pi^2 \over 720\zeta^4} \,.
\label{f'inf}
	\end{eqnarray}
This is the correct result for the boundary conditions explained
above. As remarked in the Introduction, this result is applicable for
the electromagnetic field which has the same number of degrees of
freedom as the complex scalar field. Indeed, this result agrees with
Eq.~(\ref{rhoresult}), since the free energy density in this case is
equal to the internal energy density.

\section{The temperature dependence}
We now consider the temperature dependent part of the generating
functional, which was given by the second term in
Eq.~(\ref{2terms}). Holstein and Pal
\cite{HPunpub} suggested that this 
term can be easily tackled if we rewrite it in the form
	\begin{eqnarray}
f'_\beta &=& 2 
\int {[d^3k] \over (2\pi)^3} \; \int d\omega_k \; 
\sum_{r=1}^\infty e^{-r\beta\omega_k}
\nonumber \\*
&=& -\, \sum_{r=1}^\infty {2\over r\beta} \int {[d^3k] \over (2\pi)^3} \; 
e^{-r\beta\omega_k} \,.
	\end{eqnarray}
For infinite 3-dimensional space, the integral can directly be written
in the form 
of a modified Bessel function \cite{HPunpub}. To modify this procedure
for the Casimir geometry, we first note that 
the exponent, written in terms of $\vec k$, has a square root
owing to the definition of $\omega_k$ given in
Eq.~(\ref{omegak}). This can be eliminated by using the identity 
	\begin{eqnarray}
e^{-z} = {1\over \sqrt\pi} \int_0^\infty dx\, x^{-1/2} \exp \left( -x
- {z^2\over 4x} \right) \,.
	\end{eqnarray}
Using this, for the Casimir geometry, we can utilize Eq.~(\ref{d3k})
to write
	\begin{eqnarray}
f'_\beta = -\, \sum_{r=1}^\infty {1\over \sqrt{\pi r^2\beta^2} \, \zeta}
\int_0^\infty dx\, x^{-1/2} e^{-x}
\sum_{l=-\infty}^\infty \int {d^2k_\perp \over (2\pi)^2} \exp \left( 
- {r^2\beta^2\omega_k^2 \over 4x} \right) \,.
	\end{eqnarray}
Using 
	\begin{eqnarray}
\omega_k^2 = k_\perp^2 + (\pi l/\zeta)^2 + M^2
	\end{eqnarray}
as is appropriate for this case, we can perform the integration over
$k_\perp$ in a straight forward manner. This gives
	\begin{eqnarray}
f'_\beta &=& -\, \sum_{r=1}^\infty {1\over (\pi r^2\beta^2)^{3/2} \,\zeta}
\int_0^\infty dx\, x^{+1/2} 
\exp \left( -x - {r^2\beta^2 M^2 \over 4x} \right) 
\sum_{l=-\infty}^\infty
\exp \left( {-r^2\beta^2 \pi^2 l^2 \over 4x\zeta^2} \right) 
\nonumber\\ 
&=& -\, \sum_{r=1}^\infty {1\over (\pi r^2\beta^2)^{3/2} \,\zeta}
\int_0^\infty dx\, x^{+1/2}  
\exp \left( -x - {r^2\beta^2 M^2 \over 4x} \right) 
\vartheta_3 \left(0, {i\pi
r^2\beta^2 \over 4x\zeta^2} \right) \,.
	\end{eqnarray}
Using the property of the $\vartheta$-function given in
Eq.~(\ref{thetaidentity}), this can be rewritten in the form
	\begin{eqnarray}
f'_\beta  
&=& -\, \sum_{r=1}^\infty {2\over (\pi r^2\beta^2)^2}
\int_0^\infty dx\, x  
\exp \left( -x - {r^2\beta^2 M^2 \over 4x} \right) 
\vartheta_3 \left(0, {4ix\zeta^2 
\over \pi r^2\beta^2} \right) \nonumber\\ 
&=& -\, \sum_{r=1}^\infty {2\over (\pi r^2\beta^2)^2} 
\sum_{l=-\infty}^\infty 
\int_0^\infty dx\, x 
\exp \left( -x A_{rl}^2 - {r^2\beta^2 M^2 \over 4x} \right) \,,
	\end{eqnarray}
where, for the sake of notational simplicity, we have written
	\begin{eqnarray}
1+ {4\zeta^2 
\over r^2\beta^2} l^2 \equiv A_{rl}^2 \,.
	\end{eqnarray}
The integral over $x$ can be represented in terms of the modified
Bessel function defined in Eq.~(\ref{modBessel}), and we obtain
	\begin{eqnarray}
f'_\beta = -\, \sum_{r=1}^\infty {M^2 \over (\pi r \beta)^2} 
\sum_{l=-\infty}^\infty {1\over A_{rl}^2} K_{-2}(r\beta M A_{rl}) \,.
	\end{eqnarray}
{}For arbitrary values of $M$, this cannot be reduced further
analytically. However, we are interested in the limit $M\to 0$. From
the general expression for modified Bessel functions given in
Eq.~(\ref{modBessel}), one can check that
$K_\nu(z)=K_{-\nu}(z)$. Thus, we can write $K_2$ in place of $K_{-2}$
in the last equation. Using now the limiting form of this 
function for small arguments which was given in
Eq.~(\ref{K2smallz}), we obtain
	\begin{eqnarray}
f'_\beta 
&=& -\, \sum_{r=1}^\infty {1\over 8\pi^2\zeta^4} 
\sum_{l=-\infty}^\infty {1 \over [l^2 + (r\beta/2\zeta)^2]^2} \,.
	\end{eqnarray}
The sum over $l$ can be performed by using the formula
	\begin{eqnarray}
\sum_{l=-\infty}^\infty {1\over [l^2 + a^2]^2} = {\pi\over 2a^3}
\left[ \coth \pi a + \pi a \; {\rm csch}^2\, \pi a \right] \,,
\label{sqsum}
	\end{eqnarray}
which can be obtained by differentiating Eq.~(\ref{sum}) with respect
to the parameter $a$. Using this, we obtain 
	\begin{eqnarray}
f'_\beta 
&=& -\, \sum_{r=1}^\infty {1 \over (\pi r^2\beta^2)^2} 
\left[ {\pi r\beta \over 2\zeta} \coth {\pi r\beta \over 2\zeta} +
\left( {\pi r\beta \over 2\zeta} \right)^2 \; {\rm csch}^2 \, {\pi
r\beta \over 2\zeta}  \right]\,.
\label{f'beta}
	\end{eqnarray}
This result was obtained by various authors earlier, using different
techniques~\cite{Meh67,BM69,RT89,PMG}.

One interesting check of this result is that, for $\zeta\to\infty$,
the expression in the square bracket has a limiting value of 2, so
that 
	\begin{eqnarray}
f'_\beta (\zeta\to\infty)
= -\, \sum_{r=1}^\infty {2 \over (\pi r^2\beta^2)^2} 
= -\, {\pi^2 \over 45\beta^4} \,.
	\end{eqnarray}
The internal energy density can be calculated from here, which gives
	\begin{eqnarray}
\rho'_\beta = f'_\beta + \beta {\partial f'_\beta \over \partial
\beta} = {\pi^2 \over 15\beta^4} \,,
\label{Planckrho}
	\end{eqnarray}
which is the well-known result for the energy density of a Planck
distribution. 

Experimentally, however, we are interested about small values of
$\zeta$, for which the remaining sum in Eq.~(\ref{f'beta}) has to be
performed numerically. The results are summarized in the next section.

\section{Discussion of the results}
Adding up the contributions given in Eqs.~(\ref{f'inf}) and
(\ref{f'beta}), we can write the total free energy density in the form
	\begin{eqnarray}
f_\beta = -\, {\pi^2 \over 720 \zeta^4} G (z) \,,
	\end{eqnarray}
where $z$ is a dimensionless parameter defined as
	\begin{eqnarray}
z = {2\zeta \over \pi\beta} \,,
\label{z}
	\end{eqnarray}
and
	\begin{eqnarray}
G (z) \equiv 1 + 45\sum_{r=1}^\infty {z^4 \over r^4} \left[
{r \over z} \coth {r \over z} + \left({r \over z} \right)^2 \; {\rm
csch}^2 \, {r \over z} \right] \,. 
	\end{eqnarray}
The corresponding density of the total internal energy will be given by 
	\begin{eqnarray}
\rho_\beta \equiv -\, {\pi^2 \over 720 \zeta^4} K(z) \,,
\label{defK}
	\end{eqnarray}
where
	\begin{eqnarray}
K(z) = G (z) - zG'(z) \,,
	\end{eqnarray}
the prime on the function $G$ implying differentiation with
respect to its argument. This gives
	\begin{eqnarray}
K(z) = 1 - 90
\sum_{r=1}^\infty {z^4 \over r^4} \left\{
{r\over z }\coth {r\over z} 
+ \left( {r\over z} \right)^2 \; {\rm csch}^2 \, {r\over z} 
+  \left( {r\over z} \right)^3 \; {\rm csch}^2 \, {r\over
z} \coth {r\over
z} \right\} \,.
	\end{eqnarray}

Experimentally what is measured is the force on the condenser plates
which define the boundaries of the region in the $z$-direction. The
force per unit area, or the pressure $p$, is related to the total free
energy $F$ of a system through the thermodynamic relation
	\begin{eqnarray}
p = - \left( {\partial F \over \partial V} \right)_\beta = - f_\beta -
V \left( {\partial f_\beta \over \partial V} \right)_\beta \,,
	\end{eqnarray}
where $V$ is the volume of the region, and $\beta$, or the
temperature, is to be kept constant while taking the partial
derivative. In the present situation, we can write this as
	\begin{eqnarray}
p = -f_\beta - \zeta
\left( {\partial f_\beta \over \partial \zeta} \right)_\beta \,.
	\end{eqnarray}
Substituting the expression for $f_\beta$ obtained above, we obtain
	\begin{eqnarray}
p = -\, {\pi^2 \over 240 \zeta^4} H(z) \,,
\label{defH}
	\end{eqnarray}
where $z$ is the variable defined in Eq.~(\ref{z}), and 
	\begin{eqnarray}
H(z) &=& G(z) - {1\over 3} z G'(z) \nonumber\\*
&=&  
1 - 30 \sum_{r=1}^\infty  {z\over r}
\; {\rm csch}^2 \, 
{r\over z} \; \coth \, {r\over z} 
	\end{eqnarray}
Notice again that, in the limit $\zeta\to\infty$, this gives the
familiar formula for the radiation pressure, which is one-third the
internal energy density given in Eq.~(\ref{Planckrho}).

\begin{figure}
\centerline{\epsfxsize=0.7\textwidth \epsfysize=0.3\textheight
\epsfbox{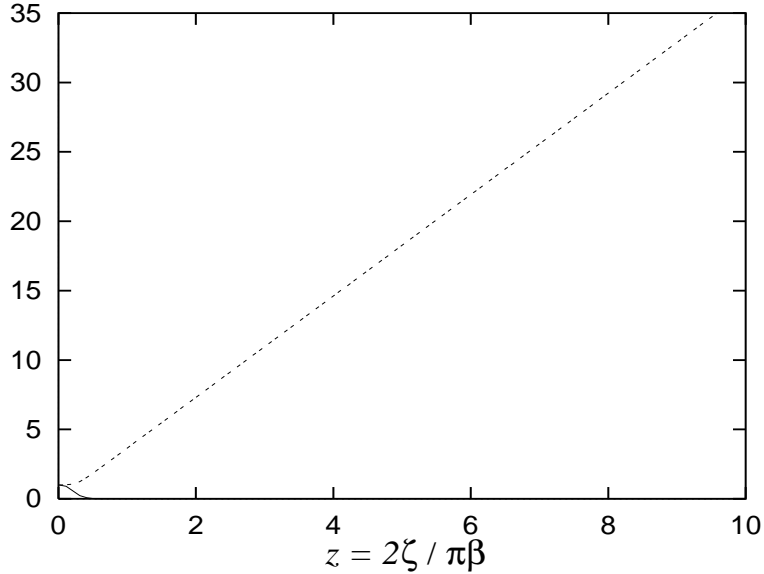}} 

\caption[]{\sf The solid line is a plot of the function $\widetilde
K(z)$, and the dotted line is of $\widetilde H(z)$. These functions
give the difference between the internal energy density and pressure
of the inside and outside regions. Both are normalized to have the
value unity at zero temperature.}\label{plots}

\end{figure}
The expressions for internal energy and pressure given above are valid
if the region inside the capacitor plates are at a temperature
$1/\beta$ and the outside region is at zero temperature. In an
experiment, this is a difficult situation to realize, specially since
the distance $\zeta$ between the plates is very small. Rather, a more
practical scenario is when the region between the plates, as well as
the ambient region, are both kept at a temperature $1/\beta$. In this
case, the quantities of physical interest are the differences in the
internal energy and pressure between the two regions --- the one
inside the plates, and the one outside. 
We denote these quantities by essentially the same notation used in
Eqs.\ (\ref{defK}) and (\ref{defH}, with two new functions
$\widetilde K(z)$ and $\widetilde H(z)$ replacing $K(z)$ and $H(z)$
respectively. These new functions are given by
	\begin{eqnarray}
\widetilde K(z)
=  1 - 90
\sum_{r=1}^\infty {z^4 \over r^4} \left\{
{r\over z }\coth {r\over z} 
+ \left( {r\over z} \right)^2 \; {\rm csch}^2 \, {r\over z} 
+  \left( {r\over z} \right)^3 \; {\rm csch}^2 \, {r\over z} 
\coth {r\over z} -3 \right\}  \,,
	\end{eqnarray}
and
	\begin{eqnarray}
\widetilde H(z) = 1 - 30 \sum_{r=1}^\infty \left[ 
{z\over r} \; {\rm csch}^2 \, 
{r\over z} \; \coth \, {r\over z} - \left( {z\over r}
\right)^4  \right]\,.
	\end{eqnarray}
In Fig.~\ref{plots}, we show these functions $\widetilde K(z)$ and
$\widetilde H(z)$ as functions of the dimensionless
variable $z$. The left end of the plots correspond to small $z$, i.e.,
large $\beta$ or vanishing temperature. In this case, the results
should reduce to the ordinary Casimir results, so that the gains
should equal unity, as seen in the figure. As $z$ increases, we see
that the energy density quickly approaches the result appropriate for
infinite volume.

As far as the difference between the external and internal pressures
is concerned, we see that the relevant function $\widetilde H(z)$
increases roughly linearly with $z$ when $z$ is large. Now, $z$ can
increase either because $\zeta$ increases, or because the temperature
increases. In the case when $\zeta$ grows, the pressure difference,
despite the increase of the function $\widetilde H(z)$, decreases
owing to the factor of $1/\zeta^4$ in Eq.~(\ref{defH}), and vanishes
in the infinite volume limit as expected. On the other hand, if we
keep $\zeta$ fixed and increase the temperature, we see that although
the difference in energy density between the external and the internal
regions goes to zero, the pressure difference in fact increases with
the temperature. This increase in the pressure difference is actually
offset by an increase in entropy difference to keep the energy
difference zero.

\paragraph*{Acknowledgments~: }
We thank P. Majumdar for discussions and S. Sinha for enlightening us
on various aspects of earlier work done in the field. After a first
version of this paper was written and submitted to the electronic
archive, S.\ Sinha, S.\ Odintsov, and F.\ 
Ravndal have brought various earlier references to our attention. We
thank all of them. AKG wants to thank the hospitality of the Saha
Institute of Nuclear Physics where this work was performed.

\end{document}